\documentstyle[12pt]{article}

\title{ \vspace*{2.5cm} \bf \LARGE 
	Study of the nature of $\iota/\eta(1440)$ \smallskip\\
        in the chiral perturbation theory approach \vspace*{5 mm}}
	
\author{ M. L. Nekrasov \thanks{ E-mail: {\tt nekrasov@mx.ihep.su}}
\smallskip  \\
{\small\it Institute for High Energy Physics, 142284 Protvino, Russia }}

\date{}

\begin{document}

\maketitle

\begin{abstract}

The nature of $\iota/\eta(1440)$ is analysed in the framework of
the hypothesis that it represents a single pseudoscalar
resonance. Assuming that it arises due to the mixing between the
glueball and the $q\bar q$ nearby states ($\eta$, $\eta'$, and
their radial excitations) two upper estimates are obtained for
the partial width $\Gamma(\iota/\eta(1440) \rightarrow K^* K)$
--- one for the case when $\iota/\eta(1440)$ is mainly a
glueball and another one when it is mainly a radial excitation
of the $s\bar s$ state. Both estimates are obtained in the
chiral perturbation theory approach taking into account the
available data on the vector mesons and the pseudoscalar state
$K(1460)$, which is interpreted as a radial excitation of the $K$
meson. The same partial width is independently estimated on the
basis of the combined OBELIX and Crystal Barrel data on the
production of $\iota/\eta(1440)$ in $p\bar p$ annihilation.
Comparing the results we show that the glueball content of
$\iota/\eta(1440)$ is suppressed while its $s\bar s$
radial-excitation interpretation is favoured by the data.

\end{abstract}

\newpage

\section{ Introduction }

The pseudoscalar (PS) state $\iota/\eta(1440)$ is traditionally
considered as a probable candidate for glueballs. It was first
discovered in reaction of $p\bar p$ annihilation in the middle
of 1960-s \cite{Baillon}. Then it has been studied in a lot of
works, both theoretical and experimental \cite{PDG}. The
hypothesis of the glueball origin of $\iota/\eta(1440)$ is
based, mainly, on the fact that it is copiously produced in
gluon-rich reactions, such as the $J/\psi$ radiative decays and
the $p\bar p$ annihilation. In addition, the glueball origin of
$\iota/\eta(1440)$ is corroborated by the fact that it is seen
in various decay modes allowed by strong interactions but is
almost invisible in $\gamma\gamma$ collisions.

However the glueball origin of $\iota/\eta(1440)$ has never been
proved, since the above arguments remain rather qualitative.
Moreover, in the last years there arose a serious doubt that
$\iota/\eta(1440)$ is really a glueball. The doubt is caused
mainly by the results of lattice calculations \cite{Bali} which
predict the lowest PS glueball with appreciably higher mass than
the observed mass of $\iota/\eta(1440)$. Simultaneously, the
experimental situation changed because there appeared some new
data which indicated that there might be two overlapping PS
resonances in the $\iota/\eta(1440)$ region. (This question,
however, is not quite clear yet \cite{PDG}.) If the latter
result will be confirmed then the $\iota/\eta(1440)$ problem
will become more intricate.

A fresh view on the problem of $\iota/\eta(1440)$ has been
recently proposed in \cite{B-Z}. In this work $\iota/\eta(1440)$
is considered as a single PS resonance whose nature is
attributed to the $s\bar s$ radial excitation. Allowing its
mixing with the higher mass PS glueball, Ref.~\cite{B-Z}
describes the Mark~III data on the production of
$\iota/\eta(1440)$ in the $J/\psi$ radiative decays. According
to \cite{B-Z}, a discrepancy with the experimental works which
prefer the two-resonance structure of $\iota/\eta(1440)$ might
be due to their use of not quite correct form of the
relativistic Breit-Wigner amplitude (the energy-dependence of
the decay widths of resonances and of some factors was not taken
into account, cf. \cite{Achasov}).

The most significant result of Ref. \cite{B-Z} is, apparently,
that it suggests a way to eliminate the discrepancy between the
experimental results which indicate the glueball origin of
$\iota/\eta(1440)$, and the results of the lattice calculations
which predict higher masses for the PS glueball. However, the
hypothesis of the $s\bar s$ origin of $\iota/\eta(1440)$ needs
to be further confirmed since not all available data on
$\iota/\eta(1440)$ have been taken into consideration in the
framework of this hypothesis. Moreover, the very discrepancy
mentioned above may turn out to be nonexistent, since the modern
lattice calculations \cite{Bali} in reality are not
model-independent as far as the $0^{-+}$ glueball is
concerned.\footnote{ The point is that there are two different
operators of the $0^{-+}$ glueball in the lattice approach. One
of them is defined as a set of three-dimensional loops deformed
in some special way (in order to produce the $0^{-+}$ quantum
numbers) \cite{Bali}. Another operator is a strictly
four-dimensional object, since it is defined as the lattice
analog of the continuous operator $G_{\mu\nu}\widetilde
G^{\mu\nu}$ \cite{TopCharge}. (The structure of the first one is
{\bf BBB} while the structure of the second one is {\bf EB}.)
Therefore, they can generate quite different glueball states.
Ref. \cite{Bali} used only the first of these two operators. One
can suppose that it generates the heavier state (which is
presumably the pseudoscalar excitation over the ground-state
scalar glueball state) whereas the second operator generates the
lighter state.} So, the glueball origin of $\iota/\eta(1440)$
cannot be completely excluded.

In the present work we carry out further investigation of
$\iota/\eta(1440)$ under the assumption of its one-resonance
structure. However, in contrast to Ref. \cite{B-Z}, we use
another set of data, namely, the data of OBELIX and Crystal
Barrel at LEAR on the $p\bar p$ annihilation at rest. Moreover,
we consider both the hypotheses on the origin of
$\iota/\eta(1440)$ --- the one according to which
$\iota/\eta(1440)$ is mainly a glueball, and another one
according to which it is mainly a radial excitation of the
$s\bar s$ state. Then both these hypotheses are to be compared
in the framework of the same approach. An intermediate case,
when $\iota/\eta(1440)$ involves comparable contributions of the
glueball and the excited $s\bar s$ state, can hardly take place,
since then the mixing partner of $\iota/\eta(1440)$ should be
visible in the gluon-rich reactions, but it is not the case if
$\iota/\eta(1440)$ is a single resonance.

The main idea of the present study is to compare the theoretical
estimate of the partial width $\Gamma (\iota \rightarrow K^* K)$
with its experimental value. (We designate further
$\iota/\eta(1440)$ by a single symbol $\iota$.) The choice of
the decay $\iota \rightarrow K^* K$ is caused by a possibility
of its description. Indeed, since this decay occurs near the
threshold its final states have small kinetic energies (in the
rest frame of $\iota$). So, the decay $\iota \rightarrow K^*K$
may be described in the framework of the chiral perturbation
theory ($\chi$PT), which is a model-independent method. The
experimental estimate of $\Gamma (\iota \rightarrow K^* K)$ may
be obtained with great accuracy as well, since from the LEAR
data it may be obtained without taking into account the
contributions of the $\rho\rho$ and $\omega\omega$ channels
(which are little-known) to the production of $\iota$. (See
Section 6 for a detailed discussion of this point.)

The structure of the present work is as follows. In the next
Section we propose a chiral effective Lagrangian which describes
PS $q\bar q$ resonances and the PS glueball. In Sec.~3 the
vector mesons are added and the vertices of the decays of PS
states to $K^*K$ are discussed. Sec. 4 shows that the chiral
loops do not change the results obtained in the previous
sections. In Sec.~5 we obtain the upper bound of $\Gamma(\iota
\rightarrow K^*K)$ while taking into account the mixing of the
PS states and the effect of the finite width of the $K^*$ meson.
In Sec.~6 the experimental value of $\Gamma(\iota\rightarrow
K^*K)$ is estimated from the combined data of OBELIX
\cite{OBELIX1,OBELIX2} and Crystal Barrel \cite{CrBr}. Sec.~7
discusses the results. Appendix~A collects the formulae which
permit to calculate the correction factors caused by the finite
widths of intermediate resonances. Appendix B estimates the
contribution of the decay $\iota \rightarrow \rho\rho$ to the
annihilation $p\bar p \rightarrow \pi\pi\iota$ at rest. Appendix
C discusses the amplitude of the $p\bar p \rightarrow
\pi\pi\iota$ in $\chi$PT.

\section{ Excited $q\bar q$ states and PS glueball in $\chi$PT }

In order to define $\chi$PT the approach of chiral effective
Lagrangian is conventionally employed \cite{Weinberg1,G-L}. The
fundamental ingredients of this approach are the interpolating
fields of observable states involved in the process to be
described. The range of application of $\chi$PT is bounded by
the condition of low momenta of the initial and final states in
the center-of-mass frame (usually each three-momentum is
required to be much less than the $\rho$ meson mass).

Independently of the kind of the process the octet of the
lightest PS states ($\pi,K,\eta$) must be represented in the
chiral effective Lagrangian. Since their Goldstone nature the
interpolating fields of these states may be collected in a
unitary unimodular matrix $u(\phi)$ which takes values in the
coset space SU(3)$_L^{}\times$SU(3)$_R^{}/$SU(3)$_V^{}$
\cite{Sigma}.  Here SU(3)$_L^{}\times$SU(3)$_R^{}$ is the
Lagrangian chiral-group symmetry and SU(3)$_V^{}$ is the
symmetry of the vacuum in QCD. Under the chiral group $u(\phi)$
transforms non-linearly,
\begin{equation}%1
u(\phi) \rightarrow 
g^{}_{L} \, u(\phi) \: h_{}^{\dagger}(g^{}_{L},g^{}_{R},\phi) 
\,= \,
h(g^{}_{L},g^{}_{R},\phi) \: u(\phi) \: g^{\dagger}_{R} ,
\end{equation}
with $g^{}_{L,R} \in$ SU(3)$_{L,R}^{}$, and $h$ is the
compensating SU(3)$_V^{}$ transformation. In case of the
diagonal transformations, $g^{}_{L} = g^{}_{R} = g^{}_{V}$,
$h$ equals $g^{}_{V}$ and, so, $h$ becomes independent on
$\phi$. Usually $u(\phi)$ is considered in the exponential
parameterization,
\begin{equation}%2
u(\phi) = \exp\{i\phi/F\}, \quad 
\phi = \sum_{a=1,\dots,8} \phi^a \lambda^a/2 ,
\end{equation}
with $\phi^a$ and $\lambda^a$ are the interpolating fields and
the Gell-Mann matrices, $F$ is the universal octet decay
constant.

The singlet member of the nonet of the lowest PS states
($\eta'$) must be described as a heavy state since it is not a
Goldstone boson. Being singlet its interpolating field,
$\phi^0$, is invariant under the chiral group. However, $\phi^0$
is not invariant under the chiral U(1)$_A$ rotation
\cite{Witten}:
\begin{equation}%3
\phi^0 \rightarrow \phi^0 + F_0\omega^0_5.
\end{equation}
Here $\omega^0_5$ is the parameter of U(1)$_A$, $F_0$ is a
dimensional constant. The nature of transformation (3) is
considered in detail in \cite{N}. Here we notice only that it is
the exceptional property of $\phi^0$ because all other
interpolating fields are invariant under U(1)$_A$.

Other heavy interpolating fields, if they are not singlets, are
not invariant under the chiral group. For instance, the octet
heavy states transform like as follows \cite{Sigma,E1}:
\begin{equation}%4
R \rightarrow h\,R\,h^{\dagger}, \quad
R = \sum_{a=1,\dots,8} R^a \lambda^a /2 .
\end{equation}
Here $h$ is the same as in (1). The singlet members of the
nonets ($R^0$) and other SU(3)-singlets (glueballs, for
instance) are invariant under the chiral group. So, the chiral
symmetry is not sufficient to distinguish between different
singlet states, and additional ideas are needed to do that. To
that end we shall follow the ideas of \cite{N} (see below).

Excluding the singlet-state problem, the transformation
properties of the interpolating fields determine the structure
of the chiral effective Lagrangian. In the framework of $\chi$PT
the Lagrangian is represented in form of the expansion in the
derivatives of fields and the current quark masses. The terms
without the derivatives are responsible for the mass spectrum of
the observable states. In case when the effective theory is to
describe the ground-state PS mesons ($\phi^0$, $\phi^a$), their
radial excitations ($P^0$, $P^a$), and the ground-state PS
glueball ($G$), the mass terms at order $p^0 + p^2$ are
determined by the following chiral-invariant Lagrangian:
\begin{eqnarray}%5
& L^{\mbox{\scriptsize {mass}}} \; = \;
  - \,\frac{1}{2}\, A_0 \,(P^0)^2 - \,A\,\langle P^2 \rangle
  - \,\frac{1}{2} M_0^2 (\phi^0)^2 - \,\frac{1}{2} M_G^2 G^2 
  - \,q \, \phi^0 G                     
  + \,\frac{F^2}{4}\langle \chi_{+}\rangle                     &
\nonumber\\[0.5\baselineskip]
& - \,\tilde\alpha_0 (\frac{\lambda^0}{2})^2 
                      \langle P^0 P^0\,\chi_{+}\rangle
  - \,\alpha_0 \frac{\lambda^0}{2} \langle P^0 P\,\chi_{+}\rangle
  - \,\alpha\, \langle PP\,\chi_{+}\rangle
  + \,i \frac{F}{2} \beta_0 \frac{\lambda^0}{2}
                            \langle P^0\,\chi_{-}\rangle  
  + \,i \frac{F}{2} \beta   \langle P\,\chi_{-}\rangle         &
\nonumber\\[0.5\baselineskip]
& + \,i \frac{F}{2} \tilde\beta_0\frac{\lambda^0}{2}
                            \langle \phi^0\,\chi_{-}\rangle 
  + \,\tilde\gamma_0 (\frac{\lambda^0}{2})^2
                            \langle \phi^0 \phi^0\,\chi_{+}\rangle 
  + \,\gamma_0 (\frac{\lambda^0}{2})^2
                            \langle \phi^0 P^0\,\chi_{+}\rangle 
  + \,\gamma    \frac{\lambda^0}{2}
                            \langle \phi^0 P\,\chi_{+}\rangle . &
\end{eqnarray}
(Notice, the particular expression (5) for the Lagrangian is not
U(1)$_A$ invariant, but it may be made invariant by means of
replacing $\phi^0$ to the U(1)$_A$-invariant combination $\phi^0
+ F_0 \Theta$ with $\Theta$ is a source of the gluon anomaly
operator in QCD \cite{G-L,Witten}.) In formula (5) the brackets
$\langle \cdots \rangle$ mean the trace operation, $\lambda^0 =
\sqrt{2/3}$. Parameters $A_0$, $A$, $M_0$, and $M_G$ describe
the masses of $P^0$, $P^a$, $\phi^0$, and $G$, respectively.
Parameter $q$ describes $\phi^0-G$ mixing in the chiral limit
(the limit of the massless quarks and the switched-off mass-like
external field). The linear in the current quark masses
contributions are described by parameters
$\alpha,\beta,\gamma$'s and by quantities $\chi_{\pm} =
u^{\dagger}\chi u^{\dagger} \pm u \chi^{\dagger} u$
\cite{G-L,E1}. Here $\chi = 2BM$ with $B$ is proportional to the
condensate of quarks, and $M$ is a mass-like external field.
Simultaneously $M$ describes the contributions of the current
quark masses; when the external field is switched off, $M =
\mbox{diag} (m_u,m_d,m_s)$. With the switched-on external field
$\chi_{\pm}$ transform like $P$, providing thus the Lagrangian
with the chiral invariance. With the switched-off external field
$\chi_{\pm}$ describe the chiral symmetry breaking. In addition,
with $m_s \not= m_{u,d}$, $\chi_{\pm}$ describe the flavour
symmetry breaking.

Now let us discuss the singlet field contributions. Note, $P^0$
and $G$ are involved not symmetrically in (5). This is caused by
the theorem \cite{N} which states that any heavy singlet PS
interpolating field, which is different from $\phi^0$, may not
contribute both to terms which involve $\chi_{\pm}$ and to the
term which describes the mixing of this state with $\phi^0$ in
the chiral limit. Due to this theorem there are two alternative
ways to involve a heavy singlet PS state to the effective
theory.

In case of the glueball state $G$ we use the possibility
according to which the chiral-limit mixing between $G$ and
$\phi^0$ is allowed but contributions of both $G$ and
$\chi_{\pm}$ to the same terms are suppressed. This choice is
caused by the following reasons. First, the $\phi^0-G$ mixing
should indeed take place, so long as QCD is possessed of the
annihilation mechanism which permits transition between $q\bar
q$ singlet states and gluonic colorless states. Second, in QCD
the quark-gluon interaction does not distinguish the quark
flavours.  So, the interpolating field of the genuine glueball
should not contribute to terms which break down the flavour
symmetry. Consequently $G$ should not contribute to terms which
involve $\chi_{\pm}$. Let us note, that we could expect the
latter property to be valid not only in the next-to-leading
order but rather in the all orders of $\chi$PT. (About the
possibility to introduce the genuine-glueball interpolating
field, especially with taking into account the UV
renormalization in QCD, see \cite{N}.)

In case of the excited state $P^0$ we use another possibility
according to which the $\phi^0-P^0$ mixing is suppressed in the
chiral limit but instead of this the contribution of $P^0$ to
terms which involve $\chi_{\pm}$ is allowed. This choice is
caused by the result of the reverse assumption. Indeed, let us
assume that there is the $\phi^0-P^0$ mixing in the chiral
limit.  Then the excited state $P^0$ can transform to a
non-excited state ($\phi^0$) without emission of
strong-interacting massless particles --- the pions and kaons in
the chiral limit. However, if there is not mass (energy) gap
then such particles should necessarily be emitted in course of
any transformation of the excited state. So, the absence of the
emission contradicts to the condition that $P^0$ is the excited
state. Therefore, the above assumption is wrong. Notice,
analogously one can show that the mixing between $P^0$ and $G$
is suppressed as well. The same result follows also from the
consistency condition: after $G$ was integrated out the $P^0 -
\phi^0$ mixing in the chiral limit would not appear if there was
not the $P^0-G$ mixing.

Extracting from (5) the quadratic terms we can describe the
spectrum of the observable states. In the channel of pions and
kaons we obtain
\begin{eqnarray}%6
& L^{\mbox{\scriptsize {mass}}}_{(\pi,K)} = 
-\frac{1}{2}M_{\pi}^2 (PP)^{\pi} -\frac{1}{2}M_{K}^2 (PP)^{K} +
\beta\left[m_{\pi}^2 (P\phi)^{\pi} + 
m_K^2 (P\phi)^K\right]  & 
\nonumber\\[0.3\baselineskip]
&\!\!-\frac{1}{2}m_{\pi}^2(\phi\phi)^{\pi}
   -\frac{1}{2}m_K^2(\phi\phi)^K. &
\end{eqnarray}
Here $m_{\pi}^2$ and $m_{K}^2$ are the masses of the pions and
kaons, $M_{\pi}^2 = A + 2\alpha m_{\pi}^2$ and $M_{K}^2 = A +
2\alpha m_{K}^2$ are the masses of the excited states $P^{\pi}$
and $P^K$. (The mixings $\phi^{\pi} - P^{\pi}$ and $\phi^{K} -
P^K$, which are controlled by $\beta$, give rise to the
corrections to the masses of order $p^4$. As far as such
corrections are beyond the level of accuracy of (6), we neglect
these mixings.) Identifying $P^{\pi}$ and $P^K$ with the real
states $\pi(1300)$ and $K(1460)$ \cite{PDG,Kataev} one can
estimate the relevant parameters of the Lagrangian: $A =
(1.28\;\mbox{GeV})^2$, $\alpha = 0.49$.

In the isosinglet channel we obtain (with taking into account
$\tilde\beta_0 = 1$ \cite{N}, and assuming, for simplicity,
$\tilde\alpha_0 = (\alpha_0 + \alpha)/2$)
\begin{eqnarray}%7
&L^{\mbox{\scriptsize {mass}}}_{(0,8,G)} \; = \;
 -\frac{1}{2} M_{N}^2 (P^N P^N) - \frac{1}{2} M_{S}^2 (P^S P^S) 
  - M_{NS}^2 (P^N P^S)  &
\nonumber\\[0.3\baselineskip]
& + \,m_{\pi}^2 (P^N\widetilde\phi^N) + 
(2m_K^2 - m_{\pi}^2) (P^S\widetilde\phi^S) 
-\frac{1}{2}M_G^2 G^2 - q \, \phi^0 G &
\nonumber\\[0.3\baselineskip]
&-\frac{1}{2}\!\left(M_0^2 - 2\tilde\gamma_0
\frac{2m_K^2\!+m_{\pi}^2}{3}\right)(\phi^0)^2
-\frac{1}{2}\frac{4m_K^2\!-m_{\pi}^2}{3}(\phi^8)^2
-2\sqrt{2}\frac{m_{\pi}^2-m_K^2}{3}(\phi^0\phi^8) . & 
\end{eqnarray}
Here $M_{N}^2 = \frac{1}{3}[2A_0 + A + (2\alpha_0 + \alpha)
2m_{\pi}^2]$ and $M_{S}^2 = \frac{1}{3}[A_0 + 2A + (\alpha_0 +
2\alpha)(4m_K^2 - 2m_{\pi}^2)]$ are the masses of the excited
states $P^N = \sqrt{2/3} P^0 + \sqrt{1/3} P^8$ and $P^S =
\sqrt{1/3} P^0 - \sqrt{2/3} P^8$. Parameter $M_{NS}^2 =
(\sqrt{2}/3)[A_0 - A + (\alpha_0 - \alpha)2m_{K}^2]$ describes
their mutual mixing. The interpolating fields
$\widetilde\phi^{N}$ and $\widetilde\phi^{S}$ involve parameters
$\beta$'s and $\gamma$'s. As far as these parameters describe
the next-to-leading order of Lagrangian (5), one may neglect the
differences between these parameters because these differences
are additionally suppressed in the large-$N_c$ expansion. So,
let us put $\beta_0 = \beta$ and $\gamma_0 = \gamma$. In this
approximation $\widetilde\phi^N = \gamma \sqrt{2/3} \phi^0 +
\beta \sqrt{1/3} \phi^8$, $\widetilde\phi^S = \gamma \sqrt{1/3}
\phi^0 - \beta \sqrt{2/3} \phi^8$.

In the isosinglet channel it is too difficult to estimate the
parameters because of their multiplicity. Nevertheless, the
problem may be simplified if one identifies $P^N$ with
$\eta(1295)$ which has been only seen in the $\eta\pi\pi$
channel \cite{PDG,GAMS}. This identification is corroborated by
the phenomenological equality $M_N^2 \approx M_{\pi}^2$ and by
observation that $A_0 = A$ and $\alpha_0 = \alpha$ in the limit
of the large $N_c$. The direct consequence of this
identification is the ideal mutual mixing between the isosinglet
PS excited states ($M_{NS}^2 = 0$). In this approximation $M_S^2
= 2M_K^2 - M_{\pi}^2 \simeq (1.6 \,\mbox{GeV})^2$.

\section{ Vector mesons in $\chi$PT }

Now let us involve the vector mesons ($\rho,\omega,\varphi,K^*$)
and discuss their interactions with the PS mesons. Namely, we
shall be interested in the vertices of the kind $V\phi\phi$,
$VP\phi$ and $VG\phi$. To the purpose there will be needful the
following auxiliary quantities composed on the interpolating
fields of the light PS mesons:
\begin{equation}%8
\Gamma_{\mu} = 
  \frac{1}{2}\left(u^{\dagger}\partial_{\mu}u +
  u\partial_{\mu}u^{\dagger}\right), \quad
u_{\mu} = 
  \frac{i}{2}\left(u^{\dagger}\partial_{\mu}u -
  u\partial_{\mu}u^{\dagger}\right). 
\end{equation}
Since $\Gamma_{\mu}$ transforms inhomogeneously under the chiral
group it allows one to define the covariant derivative of the
heavy fields:
\begin{equation}%9
\nabla_{\mu}R = \partial_{\mu}R + [\Gamma_{\mu},R].
\end{equation}
Quantity $u_{\mu}$ transforms homogeneously, so it is simply a
vector-like building block.

The leading-order chiral effective Lagrangian which describes
$V\phi\phi$ interaction, and which is chiral invariant, p- and
c-parity even, is as follows \cite{E2}:
\begin{equation}%10
L_{V\phi\phi} = 
  -\,ig \langle V_{\mu\nu} [u_{\mu},u_{\nu}] \rangle  
  -\,ig'\langle V_{\mu} [u_{\mu},\chi_{-}] \rangle  .
\end{equation}
Here $V_{\mu}$ is the vector-meson interpolating field,
$V_{\mu\nu} = \nabla_{\mu}V_{\nu} - \nabla_{\nu}V_{\mu}$ is the
tensor of the vector field. Notice, in spite of the ``naive''
chiral counting rules which require the chiral dimension of
Lagrangian (10) to be 3, the true leading term of the Lagrangian
is of order $p^1$. Indeed, the first term in (10) may be
represented in form of the expansion
\begin{equation}%11
-ig \langle V_{\mu\nu} [u_{\mu},u_{\nu}] \rangle  \, = -ig/F^2
\langle V_{\mu\nu} [\partial_{\mu}\phi,\partial_{\nu}\phi] \rangle
 + \cdots .
\end{equation}
Here the ellipsis means multi-$\phi$ contributions. Transferring
one derivative from $[\partial_{\mu}\phi,\partial_{\nu}\phi]$ to
$V_{\mu\nu}$ and taking into account the equation of motion
$\partial_{\mu}V_{\mu\nu} = - M_V^2 V_{\nu} +\cdots$, one can
reduce the number of derivatives in (11). As a result the true
leading term of Lagrangian (10) is
\begin{equation}%12
L^{\mbox{\scriptsize (vertex)}}_{V\phi\phi} = -2ig^{}_{V\phi\phi}
\langle V_{\mu} [ \phi,\partial_{\mu}\phi] \rangle .
\end{equation}
Here $g^{}_{V\phi\phi}$ means the low-energy coupling constant.
The chiral corrections to (12) begin with order $p^3$.

The chiral-invariant, p- and c-parity even Lagrangian which is
responsible for $VP\phi$ interaction at order $p^1+p^3$ is as
follows: 
\begin{eqnarray}%13
& L_{P\phi\phi} \; = \; 
   -\,ig^{}_1\langle V_{\mu} [ P,u_{\mu}] \rangle  
   -\,ig^{}_2 \langle V_{\mu\nu} [\nabla_{\mu}P,u_{\nu}] \rangle 
   -\,ig^{}_3 \langle V_{\mu} [\nabla_{\mu}P,\chi_{-}] \rangle    &
\nonumber\\[0.3\baselineskip]
&  -\,ig^{}_4 \langle V_{\mu} \{\chi_{+},[P,u_{\mu}]\} \rangle    
   -\,ig^{}_5 \langle V_{\mu} \{u_{\mu},[P,\chi_{+}]\} \rangle   
   -\,ig^{}_6 \langle V_{\mu} \{P,[\chi_{+},u_{\mu}]\} \rangle .  &
\end{eqnarray}
Here under $P$ we understand the sum of the octet fields $P^a$
multiplied by $\lambda^{a}/2$, and simultaneously the singlet
fields $P^0$ and $\phi^0$ (the singlet fields may contribute
with their own coupling constants). Let us notice, that in (13)
only the first term of the expansion $\chi_{+} = 4BM + \cdots$
is relevant, since all other terms of the expansion are
responsible for the higher vertices which involve too many
pseudoscalar fields (so $\chi_{+}$ plays the role of a
``spurion''). The leading-order $VP\phi$ vertex which is implied
by (13) is as follows
\begin{equation}%14
L^{\mbox{\scriptsize (vertex)}}_{VP\phi} = -2ig^{}_{VP\phi}
\langle V_{\mu} [P,\partial_{\mu}\phi] \rangle .
\end{equation}
It has the chiral dimension 1 and the chiral corrections
beginning with order $p^3$, as well.

Let us note, that the singlet fields do not contribute to (14)
since the vanishing commutator. Actually, this property is
manifestation of the well-known selection rule \cite{Lipkin}
imposed by SU(3) symmetry for c-parity even singlet PS state
decays.  However, with the symmetry is broken this selection
rule is no longer valid. Indeed, due to the last term in (13)
the decay $P^0\rightarrow V\phi$ is possible owing to the
``spurion'' $\chi_{+}$.

The above results may be generalized for the PS glueball, as
well, but one has to remember that the genuine glueball must not
contribute to the $\chi_{\pm}$-involving terms. The corrections
caused by the higher derivatives must be suppressed, too,
because of the above selection rule \cite{Lipkin}. So, in case
of the glueball we have in any order of $\chi$PT
\begin{equation}%15
L^{\mbox{\scriptsize (vertex)}}_{VG\phi} = 0 .
\end{equation}

\section{ Loop corrections }

Strictly speaking, the above analysis may not be complete until
the chiral loop corrections are taken into consideration. It is
well-known that in case when only the light PS mesons are
involved the chiral loops contribute to order $p^{d+2}$ if they
are calculated on the basis of order $p^d$ \cite{G-L}. However,
with the heavy fields are involved this picture may change. Let
us verify whether this is the case.

To begin with the analysis let us notice that in order to derive
only the leading loop corrections one not necessarily has to
observe the mixing effects caused by the current quark masses.
So, let us neglect these mixings and retain in the Lagrangian
only the $\phi^0 - G$ mixing, which is solely the heavy-state
mixing. Then we may immediately proceed to the formalism of the
heavy static fields with fixed four-velocity $v_{\mu}$, $v^2 =
1$ \cite{Heavy,Wise}. In this formalism virtual heavy states
cannot be destroyed or created, but can transform to other heavy
states with almost the same four-velocity $v'_{\mu}$, with
$v'_{\mu} - v_{\mu} = O(p)$. A transition to this formalism is
provided by the formula $R(v;x) = \sqrt{2M} \exp \{iMvx\} R(x)$
where $M$ is a typical mass of the heavy states,\footnote{ We
suppose that the mass splitting of the heavy states is
numerically of order $O(p)$ or less. In cases of our interest
this property takes place. Really, the mass splitting between
$\iota$ and $K^*$ is of order $O(p)$, since $M_{\iota} \approx
M_{K^*}+m_K$. The mass splitting among the PS excited states and
among the vector mesons is of order $O(p^2)$ in both cases ---
see Eqs. (6), (7), and Ref. \cite{Wise}. Notice, one may neglect
such mass splittings deriving the leading loop corrections. }
and $R(v;x)$ is a low-frequency field that depends on the
four-velocity.  Since four-momenta of the heavy states are of
the form $P = Mv + k$, with $k = O(p)$, one gets $P^2 - M^2 =
2M(kv) + O(p^2)$. So, the dependence on the large mass $M$
disappears in the propagators of the heavy states, but appears
instead in denominators of the heavy state vertices.

Basing on this result one may estimate the order in the chiral
expansion of any chiral-loop diagram. In particular, the chiral
dimension $D$ of a diagram with one heavy-field line going
through the diagram is
\cite{Wienberg2}
\begin{equation}%16
D = 2L+1 + \sum_{d=2,4,\dots} (d-2) N^{(l)}_d + 
\sum_{d=1,2,\dots} (d-1) N^{(hl)}_d \ge 2L+1 .
\end{equation}
Here $L$ denotes the number of light-meson loops, $N^{(l)}_d$
($N^{(hl)}_d$) counts the number of light-meson
(heavy-and-light-meson) vertices of the chiral dimension $d$. As
applied to the vertices of the previous section this result
means that the chiral loop corrections begin with order $p^3$.
Let us emphasize that this is the same order in which the usual
chiral corrections begin with to the vertex $VP\phi$.

In case of the glueball-involving vertex $VG\phi$ one has to
carry out more detailed analysis which would take into account
the property that the current quark masses should not contribute
to the vertices which describe the glueball interactions. Let us
recall that this condition is the external one with respect to
the effective theory. So, it must be satisfied in the presence
of the chiral loops as well as in their absence. It is clear
that in the presence of the chiral loops it may be only
satisfied when there are not chiral loop corrections to the
glueball-involving vertices. Really, the chiral loops always
produce quark-mass-dependent factors like $m_q\ln m_q$
(remember, the heavy static fields do not form closed loops).
So, since the quark-mass dependence is suppressed in the
glueball-involving vertices, the chiral loops must not
contribute to them. The mechanism ensuring this effect consists
in the property that the vertices which involve both $G$ and the
light PS mesons are suppressed in the effective theory. Indeed,
if in the chiral effective Lagrangian there are bare vertices
$RG(\phi)^{n}$, $n \ge 1$, with $R$ is a heavy static field,
then via the tadpole diagrams these vertices give rise to the
quark-mass-dependent factors in the renormalized vertices
$RG(\phi)^{n-2}$. (In case with $n=1$ one might consider more
complicated diagrams which involve more than one bare vertex.)
Thus, so long as the quark-mass dependence is suppressed in the
vertices $RG(\phi)^{n-2}$, the Lagrangian vertices $RG(\phi)^n$
must be suppressed from the very beginning. As a result, the
full renormalized vertices $RG(\phi)^n$ are suppressed as
well.\footnote{ Let us note, that the above discussion concerns
one-particle-irreducible diagrams only. Concerning
one-particle-reducible diagrams, they may well describe an
interaction between the glueball and the light PS mesons, but
only through the $G-\phi^0$ mixing in the external lines
outgoing from the vertices. Moreover, such diagrams may yield an
effective quark-mass dependence in the glueball interactions
with $q\bar q$ states, but only through the external-line mixing
which occurs outside the vertices.}

Now let us discuss the chiral loop corrections to the Lagrangian
that describes the spectrum of the heavy PS states. In
accordance with (16) these corrections begin with order $p^3$.
However the more detailed analysis shows that the relevant
diagrams are the one-loop ones of the type of the self-energy
with two $VP\phi$ vertices. In the leading order such diagrams
contribute only to the kinetic terms of the heavy states, which
in the static-field formalism are of the form
$iR^{\dagger}(kv)R$. So, the $p^3$-order loop corrections
manifest themselves as $[1+O(p^2)]$-renormalization of the
heavy-state wave functions. This effect may not change the
results of Sec.~2. It is clear, also, that it does not change
the above results about the corrections to the vertices.  The
corrections that arise from the wave-function renormalization of
the light PS mesons and the vector mesons are of the same
property.

So, the above discussion shows that the chiral loops do not
change our results obtained in the quasi-classical (loop-free)
approximation. In particular, there are not chiral corrections
to the vertex $VG\phi$ which is zero. The corrections to the
vertex $VP\phi$ arise at order $p^3$ which is higher by two
units as compared with the leading order $p^1$ of this vertex.
In accordance with the current practice such corrections may be
estimated as 20\% of the leading-order result. (It should be
noted, that the individual one-loop corrections, that arise from
the vector-meson wave function renormalization, are relatively
large \cite{Wise}. Nevertheless, their flavour-non-symmetric
parts are small. So, one can redefine $\chi$PT attributing the
common large flavour-symmetric corrections to the leading-order
result --- i.e. to the flavour-symmetric coupling constants,
etc. --- and the remaining small parts of the corrections to the
proper corrections. Thus the above statement about the
20\%-estimate of the chiral corrections remains in force.)

\section{ $\Gamma(\iota \rightarrow K^* K)$ in $\chi$PT }

Assuming that $\iota$ arises due to the mixing of the pure
glueball, isoscalar lowest $q\bar q$ states, and their radial
excitations, let us present the interpolating field of $\iota$
in form of the following decomposition
\begin{equation}%17
P^{\,\iota} = {\cal O}^{\,\iota}_8 \phi^8 +
              {\cal O}^{\,\iota}_0 \phi^0 +
              {\cal O}^{\,\iota}_{\!S}  P^S +
              {\cal O}^{\,\iota}_{\!N}  P^N +
              {\cal O}^{\,\iota}_{\!G}  P^G.
\end{equation}
Here ${\cal O}^{\,n}_j$ is the orthogonal mixing matrix defined
on the basis of Lagrangian (7). Further we assume ${\cal
O}^{\,\iota}_{\!N}=0$, thinking that $P^N$ is identical with
$\eta(1295)$. Due to (12), (14), (15), and (17) the amplitude of
the decay $\iota \rightarrow K^*K$ is as follows
\begin{equation}%18
\mbox{Amp}\,(\iota \rightarrow K^*K) = \left(
\sqrt{3}\,g^{}_{V\phi\phi}{\cal O}^{\,\iota}_8 - 
\frac{1}{\sqrt{2}}\,g^{}_{VP\phi}{\cal O}^{\,\iota}_{\!S} \right)
\epsilon_{\mu}(K^{*}) p_{\mu}(K) .
\end{equation}
Here $\epsilon_{\mu}(K^{*})$ and $p_{\mu}(K)$ are the
polarization vector and the four-momentum of the $K^{*}$ and $K$
mesons. Taking into account factor 4 caused by the presence of
the two neutral and the two charge modes in the $K^*K$ system,
and taking into account the equality $\sum_{n}
{\epsilon^{(n)}_{\mu} \epsilon^{(n)}_{\nu}}p_{\mu}p_{\nu} =
|{\bf p}|^2 M_{\iota}^2 / M_{K^*}^2$ where ${\bf p}$ is the kaon
momentum in the $\iota$ rest frame, we obtain the partial width
of the decay:
\begin{equation}%19
\Gamma(\iota \rightarrow K^*K) = \frac{1}{2\pi}\left(
\sqrt{3}\,g^{}_{V\phi\phi}{\cal O}^{\,\iota}_8 - 
\frac{1}{\sqrt{2}}\,g^{}_{VP\phi}{\cal O}^{\,\iota}_{\!S} \right)^2
\frac{\xi\,|{\bf p}|^3}{M_{K^*}^2} .
\end{equation}
Here $\xi$ is a correction factor caused by the resonance
properties of the $K^*$ meson (see Appendix~1). With $M_{\iota}$
approaches the threshold $(M_{K^*}+m_K)$, $\xi$ grows rapidly,
thus compensating partly the decrease of the phase volume. In
distance of the threshold and/or neglecting the width of the
$K^*$ meson, $\xi$ approaches 1. With $M_{\iota} = 1416\pm 6$
MeV, which is the mean value of the Crystal Barrel and OBELIX
data (see below), one has $\xi = 1.56^{\,-\,0.21}_{\,+\,0.37}$.
(For comparison: with $M_{\iota} = 1440$ MeV, $\xi = 1.07$.)

Coupling constants $g^{}_{V\phi\phi}$ and $g^{}_{VP\phi}$ may be
estimated on the basis of the PDG data \cite{PDG}. Thus, from
the vector meson data one can obtain (with the leading $\chi$PT
corrections are taken into account) \cite{Bramon}
\begin{equation}%20
g_{V\phi\phi}^2/4\pi \simeq 2.9 .
\end{equation}
Constant $g^{}_{VP\phi}$ may be estimated on the basis of the
data $K(1460) \rightarrow \rho K,\,K^*\pi$ \cite{PDG}. With help
of (7) and (14) we obtain $g_{VP\phi}^2/4\pi \simeq 1.2,\,2.6$,
respectively. A noticeable difference in the results may be
explained by inaccuracy in the experimental data, and by the
fact that the final states are not enough soft in the case of
these decays (therefore, the chiral corrections may be large).
So, let us make use for the constant $g^{}_{VP\phi}$ the rough
upper bound which numerically coincides with (20) and which, we
believe, should cover the above uncertainties,
\begin{equation}%21
g_{VP\phi}^2/4\pi < 2.9 .
\end{equation}
Then we obtain the corresponding upper bound of the width 
\begin{equation}%22
\Gamma(\iota \rightarrow K^*K) \; < \; \left(
\sqrt{6}\,\vert{\cal O}^{\,\iota}_8 \vert +
\vert{\cal O}^{\,\iota}_{\!S} \vert \right)^2
\frac{2.9 \,\xi \,|{\bf p}|^3}{M_{K^*}^2} .
\end{equation}

Further we consider two cases --- the first one when $\iota$ is
mainly a glueball, and the second one when $\iota$ is mainly a
radial excitation of the $s\bar s$ state. In the first case the
mass of the excited $s\bar s$ state must be in the range or (due
to the mixing) somewhat higher than 1.6 GeV.  However no PS
state has been seen in this mass range in the gluon-rich
reactions (in the channels $K^*K$, $K\bar K\pi$). So, the
excited $s\bar s$ state can only be weakly mixed with the PS
glueball. With this property the analysis based on Lagrangian
(7) gives estimate \cite{N}
\begin{equation}%23
\vert {\cal O}^{\,\iota}_8 \vert \; \simeq \;
\frac{\sqrt{8}}{3} \, \frac{m_K^2 - m_{\pi}^2}{M_{\,\iota}^2} \,
\vert {\cal O}^{\,\iota}_0 \vert \; \simeq \;
0.1\,\vert {\cal O}^{\,\iota}_0 \vert .
\end{equation}
(Remember, there is not direct $\phi^8 - G$ mixing in the
Lagrangian, but this mixing may occur indirectly, via the
$\phi^8 - \phi^0$ and $\phi^0 - G$ mixings.)

Now, let us consider the condition of the glueball quality of
$\iota$: $\vert {\cal O}^{\,\iota}_{\!G} \vert > \vert {\cal
O}^{\,\iota}_j \vert$, $j = 8,0,S$. Together with the trivial
probability condition $\vert {\cal O}^{\,\iota}_{\!G} \vert^2 +
\sum_{j} \vert {\cal O}^{\,\iota}_j \vert^2 = 1$ it leads to
estimate $\sqrt{6}\,\vert{\cal O}^{\,\iota}_8 \vert + \vert{\cal
O}^{\,\iota}_{\!S} \vert < 0.75$. As a result, with $M_{\iota} =
1416 \pm 6$ MeV we obtain
\begin{equation}%24
\Gamma(\iota \rightarrow K^*K) \; < \; 
8.2 \pm 3.5 \;\mbox {MeV}.
\end{equation}
The error in (24) is determined as the sum (in quadratures) of
the statistical error caused by inaccuracy in $M_{\iota}$ and
the systematical error caused by $\chi$PT uncertainties in
(22) and (23), which we estimate to be 20\% in the amplitude.
Let us note, that using the more strong condition of the
glueball quality of $\iota$ one may obtain the more strong
estimate of the width.  For example, with $\vert {\cal
O}^{\,\iota}_{\!G} \vert^2 > \sum_{j} \vert {\cal O}^{\,\iota}_j
\vert^2$ one obtains
\def\theequation{24\mbox{$'$}}
\begin{equation}%24'
\Gamma(\iota \rightarrow K^*K) \; < \; 7.7 \pm 3.3 \;\mbox{MeV}.
\end{equation}

In case when $\iota$ is mainly the $s\bar s$ excited state, the
simplest way to estimate the width is to put ${\cal
O}^{\,\iota}_8 = 0$, $\vert {\cal O}^{\,\iota}_{\!S} \vert = 1$.
(Then $\Gamma(\iota \rightarrow K^*K) \simeq 14$ MeV.) However
this estimate is rather naive and cannot be realistic since the
pure $s\bar s$ excited state cannot satisfy the
$\iota$-properties. To obtain realistic estimate one must demand
\cite{B-Z} noticeable mixing between the $s\bar s$ excited state
and the higher mass PS glueball. However, in virtue of (7) the
$P^S-G$ mixing is only possible via the $P^S-\widetilde
\phi^S(\phi^0)$ and $\phi^0-G$ mixings. Therefore, the
$P^S-\phi^8$ mixing should be noticeable, too. Moreover, with
the group factor $\sqrt{6}$ the contribution of ${\cal
O}^{\,\iota}_8$ in (22) may turn out to be significant. However,
in contrast to the previous case, we cannot estimate it. So, let
us estimate the maximal upper bound of the width. It follows
from the possibly weakest conditions $\vert {\cal
O}^{\,\iota}_{\!S} \vert > \vert {\cal O}^{\,\iota}_8 \vert$,
$\vert {\cal O}^{\,\iota}_{\!S} \vert^2 + \vert {\cal
O}^{\,\iota}_8 \vert^2 < 1$, under which we obtain
\setcounter{equation}{24} 
\def\theequation{\arabic{equation}}
\begin{equation}%25
\Gamma(\iota \rightarrow K^*K) \; < \; 
87 \;\mbox {MeV}.
\end{equation}
Notice, estimate (25) is saturated when $\vert {\cal
O}^{\,\iota}_{8} \vert = \vert {\cal O}^{\,\iota}_{\!S}
\vert = 1/\sqrt{2}$. But this condition may not be real. So, the
true value of the width is, apparently, not too close to the
upper boundary indicated in (25). Unfortunately, we cannot
propose more strong estimate.

\section{ $\Gamma(\iota \rightarrow K^* K$) from LEAR data }

The experimental value of $\Gamma(\iota \rightarrow K^* K)$ has
been presented neither in the PDG \cite{PDG} nor in the original
works. In principle, one may extract it from the available data
on the $J/\psi$ radiative decays (Mark III, DM2) and the $p\bar
p$ annihilation at rest (LEAR). The more preferable between them
are the LEAR data (OBELIX + Crystal Barrel) because these data
were collected with greater statistics and, what is more
important, they permit to extract $\Gamma(\iota \rightarrow K^*
K)$ without additional assumptions. Namely, one may extract it
little knowing parameters of the (probable) decays $\iota
\rightarrow \rho\rho$ and $\iota \rightarrow \omega\omega$. The
point is that so long as these decays occur under the nominal
threshold, they may be noticeable only with the large invariant
mass of $\iota$. On the other hand, in the $p\bar p$
annihilation at rest the creation of $\iota$ with large
invariant mass is suppressed by the phase volume, which is
rapidly decreasing with the invariant mass of $\iota$ is
increasing. As a result, the contributions of $\rho\rho$ and
$\omega\omega$ to the creation of $\iota$ in the $p\bar p$
annihilation at rest turn out to be negligible (see Appendix~2).
This situation is drastically different from that which takes
place in the $J/\psi$ radiative decays, where the invariant mass
of $\iota$ is practically unlimited by the phase volume and,
therefore, the $\rho\rho$ and $\omega\omega$ contributions to
the creation of $\iota$ may turn out to be significant
\cite{Achasov}.

So, we shall use the OBELIX and Crystal Barrel data only.
Remember, OBELIX saw $\iota$ in the modes $KK\pi$ produced both
via the $K^* K$ and in the direct three-particle decays. Crystal
Barrel saw $\iota$ in the $\eta\pi\pi$ modes. Under the
assumption that $\iota$ is a single resonance OBELIX presented
its results in the framework of two fits \cite{OBELIX1}. In the
first fit there were $M_{\iota} = 1426 \pm 2$ MeV,
$\Gamma_{\iota} = 78 \pm 4$ MeV, in the second one $M_{\iota} =
1410 \pm 2$ MeV, $\Gamma_{\iota} = 56 \pm 6$ MeV.  Crystal
Barrel \cite{CrBr} obtained $M_{\iota} = 1409 \pm 3$ MeV,
$\Gamma_{\iota} = 86 \pm 10$ MeV. The statistical mean values
\cite{PDG} of these results are $M_{\iota} = 1416 \pm 6$ MeV,
$\Gamma_{\iota} = 73 \pm 4$ MeV.

Crystal Barrel \cite{CrBr} presented the absolute branching
ratio $B(p\bar p \rightarrow \pi\pi\iota,\,\iota \rightarrow
\eta\pi\pi) = (3.3 \pm 1.0)\times 10^{-3}$. This result implies
\begin{equation}%26
B(p\bar p \rightarrow \pi^{\!+}\pi^{\!-}\iota,
\,\iota \rightarrow \eta\pi\pi) = 
(2.2 \pm 0.9)\times 10^{-3}.
\end{equation}
OBELIX \cite{OBELIX2} obtained
\begin{equation}%27
B(p\bar p \rightarrow \pi^{\!+}\pi^{\!-}\iota,
\,\iota \rightarrow KK\pi) = 
(1.80 \pm 0.15)\times 10^{-3}.
\end{equation}
From (26) and (27), neglecting other possible decays of
intermediate $\iota$, there follows
\begin{equation}%28
B(p\bar p \rightarrow \pi^{\!+}\pi^{\!-}\iota) = 
(4.0 \pm 0.9)\times 10^{-3}.
\end{equation}
Analysis of the results presented by OBELIX \cite{OBELIX1}
allows one to determine the quota of $K^*K$ from the all allowed
$KK\pi$ modes:
\begin{equation}%29
\frac{B(p\bar p \rightarrow \pi^{\!+}\pi^{\!-}\iota,\,\iota
\rightarrow K^*K)}{B(p\bar p 
\rightarrow \pi^{\!+}\pi^{\!-}\iota,\,\iota
\rightarrow KK\pi)} = 0.35 \pm 0.04  .
\end{equation}
On the basis of (27)--(29) one can obtain the following
important result:
\begin{equation}%30
\frac{B(p\bar p \rightarrow \pi^{\!+}\pi^{\!-}\iota,\,\iota
\rightarrow K^*K)}{B(p\bar p \rightarrow
\pi^{\!+}\pi^{\!-}\iota)} = 0.16 \pm 0.04 .
\end{equation}

It is clear, that with the neglected resonance properties of
$\iota$ and $K^*$ the left hand size in (30) is the sought-for
branching $B(\iota \rightarrow K^* K)$. However due to the
resonance properties there may be considerable corrections. In
order to estimate them let us consider the relations
\begin{eqnarray}%31,32
&&B(p\bar p \rightarrow \pi^{\!+}\pi^{\!-}\iota,\,\iota
\rightarrow K^*K) = \xi^* B_0(p\bar p \rightarrow
\pi^{\!+}\pi^{\!-}\iota)\,B,                \\[0.5\baselineskip]
&&B(p\bar p \rightarrow \pi^{\!+}\pi^{\!-}\iota,\,\iota
\rightarrow ``other") = \bar \xi B_0(p\bar p \rightarrow
\pi^{\!+}\pi^{\!-}\iota)\,(1-B).
\end{eqnarray}
Here the single $B$ is the sought-for branching $B(\iota
\rightarrow K^*K)$, subscript ``0'' means that branching
$B_0(p\bar p \rightarrow \pi^{\!+}\pi^{\!-}\iota)$ is defined in
a speculative case of the zero widths of $\iota$ and $K^*$. The
``$other$'' in (32) means that all other decays of $\iota$ are
implied, i.e. all decays which occur not via the $K^*K$.
Quantities $\xi^*$ and $\bar \xi$ are the factors that guarantee
the equality in the relations. Summing up (31) and (32) one gets
\begin{equation}%33
B(p\bar p \rightarrow \pi^{\!+}\pi^{\!-}\iota) =
B_0(p\bar p \rightarrow \pi^{\!+}\pi^{\!-}\iota)
\left[ \xi^* B + \bar \xi (1-B) \right] .
\end{equation}
From (31) and (33) one obtains
\begin{equation}%34
\frac{B(p\bar p \rightarrow \pi^{\!+}\pi^{\!-}\iota,\,\iota
\rightarrow K^*K)}{B(p\bar p \rightarrow
\pi^{\!+}\pi^{\!-}\iota)} = 
\frac{\xi^* B}{\xi^* B + \bar \xi (1-B)} .
\end{equation}
Equating the right hand sizes in (30) and (34), and using the
property that $\xi^*$ and $\bar \xi$ are the functions on $B$
(see Appendix~1), we obtain the true value of the branching:
\begin{equation}%35
B(\iota \rightarrow K^*K) = 0.40 \pm 0.08 .
\end{equation}
Let us emphasize, that this result is more than twice as large
as the naive value in (30). Correction factors $\xi^*$ and $\bar
\xi$ turn out to be $0.60 \pm 0.01$ and $2.13 \pm 0.03$,
respectively. (Both they are far from 1, as well.) Multiplying
(35) on the total width we come to the final result
\begin{equation}%36
\Gamma^{\exp}(\iota \rightarrow K^* K) = 
29.2 \pm 6.1 \;\mbox {MeV}.
\end{equation}
This result may be compared with the theoretical estimates
(24) and (25).

\section{ Discussion and conclusions }

The main results of the present work are the theoretical
estimates (24) and (25) for the partial width $\Gamma(\iota
\rightarrow K^*K)$ --- the first one for case when $\iota$ is
mainly a glueball, and the other one for case when $\iota$ is
mainly a radial excitation of the $s\bar s$ state. (An
intermediate case, when $\iota$ involves comparable
contributions of the glueball and the excited $s\bar s$ state
is, apparently, not allowed by the data if $\iota$ is a single
resonance.) Since the above estimates are obtained in $\chi$PT
approach, their status is close to being model-independent. The
assumptions used in deriving the estimates are as follows. First
of all, we suppose that $\iota$ arises due to the mixing of the
glueball, the isoscalar lowest $q\bar q$ states ($\eta,\eta'$),
and their radial excitations. Then, we identify the excited
$n\bar n$ state with $\eta(1295)$, and suppose that $K(1460)$ is
the radial excitation of the $K$ meson. These assumptions are in
agreement with the modern understanding of the $0^{-+}$ spectrum
\cite{PDG,Kataev} and may be verified by independent methods.

Another important new result is the estimate of $\Gamma(\iota
\rightarrow K^*K)$ obtained from the combined OBELIX and Crystal
Barrel data. The idea to use specifically the OBELIX and
Crystall Barrel data is caused by the following reasons.  First,
these data were collected with the best statistics of $\iota$.
Second, and this point is more important, from kinematic reasons
the creation of $\iota$ with its subsequent decay to $\rho\rho$
and $\omega\omega$ in the $p\bar p$ annihilation at rest is
strongly suppressed. As a result, one need not take into account
these decays while extracting $\Gamma(\iota \rightarrow K^*K)$
from the data. This property essentially simplifies the analysis
based on the $p\bar p$ annihilation data as compared, for
example, with the analysis based on the $J/\psi$ radiative
decays.

An important technical point of our analysis is that it
accurately takes into account the resonance properties of
$\iota$ and $K^*$. This point is indeed important since all the
decays, considered above, occur near the threshold. We use the
relativistic Breit-Wigner amplitude \cite{Achasov} which takes
into account the dependence of the partial widths of the
resonances on their (varying) invariant masses. As a result, for
example, the true value of $B(\iota \rightarrow K^*K)$ turns out
to be more than twice as large as the ``naive'' value, which
follows directly from the data without taking into account the
resonance properties of $\iota$ and $K^*$.

Comparing the theoretical estimate (24) with the experimental
estimate (36) we conclude that with the one-resonance structure
of $\iota$ it may not be a glueball, since the ratio of the
theoretical estimate to the experimental one does not exceed
$0.28 \pm 0.13$, which is $5.5\sigma$ less than 1. However when
$\iota$ is mainly a radial excitation of the $s\bar s$ state the
theoretical estimate agrees with the experimental one. So,
taking also into account the results of Ref. \cite{B-Z}, one may
conclude that the $s\bar s$ interpretation of $\iota$ is
possible. However, in accordance with \cite{B-Z}, it is only
possible when there is noticeable mixing between the $s\bar s$
excited state and the higher mass PS glueball (with the mixing
angle about $18^{\mbox{\scriptsize o}}$). The present study
modifies this picture somewhat. Namely, we find that as soon as
$\iota$ involves a noticeable glueball contribution it must
involve also a noticeable ground-state $q\bar q$ contribution.
This (qualitative) result follows from the fact that the direct
$P^S-G$ mixing is suppressed in $\chi$PT, but it may be realized
indirectly via the $P^S-\phi^0$ and $\phi^0-G$ mixings. Of
course, to describe quantitatively this effect one has to
perform a more detailed phenomenological investigation, which
might be similar to that of \cite{B-Z} but should take into
account the results of the present study.

In conclusion, let us discuss whether our results are applicable
to the case when there are two PS resonances in the $\iota$
region.  Because the two components of $\iota$ are most likely
the PS glueball and the $s\bar s$ excited state that are
strongly mixed \cite{Close}, our estimate (24) is no longer
valid in this case since the estimate (23) becomes incorrect.
The estimate (25) remains valid, but it may be applied only to
the lower $\iota$ state. For the upper $\iota$ state, the
analogous estimate is more than 200 MeV due to increased phase
volume. So, both theoretical estimates agree with the
experimental ones which in this case may be taken directly from
the OBELIX results \cite{OBELIX1}. (Since in the case of the
two-resonance structure of $\iota$ the upper $\iota$ decays
almost only to $K^*K$, while the lower $\iota$ almost does not
decay into this channel). So, to specify the nature of the both
$\iota$'s one needs an additional investigation which must take
into account the strong mixing between the lower and upper
$\iota$ states.

\bigskip

The author is grateful for the helpful discussions to
A.B.Arbuzov, R.N.Rogalev, and P.P.Temnikov. The work was
supported in part by RFBR, grant 95-02-03704-$a$.

\section*{Appendix 1}

\setcounter{equation}{0}
\def\theequation{A1.\arabic{equation}}

This Appendix collects the formulae of calculation of the
correction factors $\xi$, $\xi^*$, and $\bar \xi$.

Let us begin with calculation of $\xi$ which is for the decay
$\iota \rightarrow K^*K$. Let $\Gamma(\iota \rightarrow K^*K;E)$
be the true partial width of $\iota$ which mass is equal to $E$,
and $\Gamma_{0}(\iota \rightarrow K^*K;M_{\iota})$ be the
speculative partial width which is taken with zero width of the
$K^*$. Then we can write
\begin{equation}%A1.1
\Gamma(\iota \rightarrow K^*K;E) = 
\xi(E)\,\Gamma_{0}(\iota \rightarrow K^*K;M_{\iota}).
\end{equation}
From (A1.1) there follows
\begin{equation}%A1.2
\xi(E) = \int^{E-m_K}_{m_K+m_{\pi}}
2E'dE'\:W(K^*;E') \left(\frac{M_{K^*}}{E'}\right)^2
\left[\frac{{\cal K}(E;E',m_K)}{
{\cal K}(M_{\iota};M_{K^*},m_K)}\right]^3.
\end{equation}
Here ${\cal K}(M;m1,m2)$ is the module of the three-momentum of
the particle $m1$ ($m2$) in the rest frame of $M$ in the decay
$M \rightarrow m1+m2$. $W(K^*;E)$ is the Breit-Wigner function,
\begin{equation}%A1.3
W(K^*;E) = \frac{1}{\pi}\,
\frac{E\,\Gamma(K^*;E)}{[M_{K^*}^2 - E^2]^2 +
[E\,\Gamma(K^*;E)]^2}.
\end{equation}
Let us emphasize, that the correct Breit-Wigner function
involves the factor $E$ before the width, and the width must be
dependent on the varying invariant mass E of the resonance
\cite{Achasov}. In case of the $K^*$ meson the latter dependence
of the width is as follows ($r=0.002$ MeV$^{-1}$)
\begin{equation}%A1.4
\Gamma(K^*;E)  =  \Gamma(K^*;M_{K^*}) \frac{M_{K^*}^2}{E^2} 
\left[\frac{{\cal K}(E;m_K,m_{\pi})}{
{\cal K}(M_{K^*};m_K,m_{\pi})}\right]^3
\frac{1+[r{\cal K}(M_{K^*};m_K,m_{\pi})]^2}{1+
[r{\cal K}(E;m_K,m_{\pi})]^2}.
\end{equation}
With help of (A1.2)--(A1.4) one can calculate $\xi(E)$. In
particular, with $E=1416\,\mbox{MeV}$ one obtains $\xi = 1.56$.

The decays $p\bar p \!\rightarrow\! \pi^{\!+}\pi^{\!-}\iota$
with $\iota \!\rightarrow\! K^*K$, and $p\bar p \!\rightarrow\!
\pi^{\!+}\pi^{\!-}\iota$ with $\iota \!\rightarrow\!``other"$
may be analysed in the similar way. So, let $\Gamma(\iota;E)$ be
the total width of $\iota$ which mass is equal to $E$.
Separating $K^*K$ from the ``$other$'' decay modes, one may
represent $\Gamma(\iota;E)$ in the form
\begin{equation}%A1.5
\Gamma(\iota;E) = \Gamma(\iota \rightarrow K^*K;E) +
\Gamma(\iota \rightarrow ``other").
\end{equation}
Here we take into account the property that $\Gamma(\iota
\rightarrow ``other";E)$ is the slowly varying function and
neglect its dependence on $E$.\footnote{ Indeed, except $\iota
\rightarrow \rho\rho$ and $\iota \rightarrow \omega\omega$, all
decays $\iota \rightarrow ``other"$ occur far from the
threshold. Concerning $\iota \rightarrow \rho\rho$ and $\iota
\rightarrow \omega\omega$, their contributions are too weak in
case of the $p\bar p$ annihilation at rest, see Appendix~2.} In
this approximation
\begin{equation}%A1.6
\Gamma(\iota \rightarrow ``other") = (1-B)\,\Gamma(\iota;M_{\iota})
\end{equation}
with $B = B(\iota \rightarrow K^*K;M_{\iota})$. The rapidly
varying function is
\begin{equation}%A1.7
\Gamma(\iota \rightarrow K^*K;E) = 
\frac{\xi(E)}{\xi(M_{\iota})}\,B\Gamma(\iota;M_{\iota}).
\end{equation}

In accordance with (A1.5) let us introduce Breit-Wigner
functions
\begin{eqnarray}%A1.8,9
&&W(\iota \rightarrow K^*K;E) \; = \; \frac{1}{\pi}\,
\frac{E\,\Gamma(\iota \rightarrow K^*K;E)}{[M_{\iota}^2 - E^2]^2 +
[E\,\Gamma(\iota;E)]^2},
\\[0.5\baselineskip]
&&W(\iota \rightarrow ``other";E) \; = \; \frac{1}{\pi}\,
\frac{E\,\Gamma(\iota \rightarrow ``other")}{[M_{\iota}^2 - E^2]^2 +
[E\,\Gamma(\iota;E)]^2}.
\end{eqnarray}
With help of these functions the correction factors $\xi^*$ and
$\bar\xi$ introduced in (31) and (32) may be represented as
\begin{eqnarray}%A1.10,11
&&\xi^* \; = \; \frac{1}{B}
\int^{M_{p\bar p}-2m_{\pi}}_{2m_K+m_{\pi}}
2EdE\:W(\iota \rightarrow K^*K;E)\:
\frac{\int d\Phi_3(E,m_{\pi},m_{\pi})}{
      \int d\Phi_3(M_{\iota},m_{\pi},m_{\pi})},
\\[0.5\baselineskip]
&&\bar \xi \; = \; \frac{1}{1\!-\!B}
\int^{M_{p\bar p}-2m_{\pi}}_{2m_K+m_{\pi}}
2EdE\:W(\iota \rightarrow ``other";E)\:
\frac{\int d\Phi_3(E,m_{\pi},m_{\pi})}{
      \int d\Phi_3(M_{\iota},m_{\pi},m_{\pi})}. \qquad 
\end{eqnarray}
Here $\int d\Phi_3(E,m_{\pi},m_{\pi})$ is the phase volume of
the annihilation $p\bar p \rightarrow E\pi\pi$ at rest, which
is corrected by the pion derivatives in the decay vertex, see
Appendix~3.

\section*{Appendix 2}

\setcounter{equation}{0}
\def\theequation{A2.\arabic{equation}}

In this Appendix we estimate the contribution of the decay
channel $\iota \rightarrow \rho\rho$ to the annihilation of
$p\bar p$-atom to $\pi\pi\iota$. Namely, we estimate quantity
$R$, where
\begin{equation}%A2.1
R = \frac{B(p\bar p \rightarrow \pi^{\!+}\pi^{\!-}\iota,\,
          \iota \rightarrow \rho\rho)}
         {B(p\bar p \rightarrow \pi^{\!+}\pi^{\!-}\iota)} .
\end{equation}
At first, let us suppose that the $\rho\rho$ contribution is
small, $R\ll 1$. (Then one may use the formulae of Appendix~1.)
Putting to use Ref. \cite{Achasov}, we can write
\begin{equation}%A2.2
R = \frac{\displaystyle
          \int^{M_{p\bar p}-2m_{\pi}}_{4m_{\pi}}
          2EdE\:W(\iota \rightarrow \rho\rho;E)\:
          \int d\Phi_3(E,m_{\pi},m_{\pi})}{\displaystyle
          \int^{M_{p\bar p}-2m_{\pi}}_{4m_{\pi}}
          2EdE\:W(\iota;E)\:
          \int d\Phi_3(E,m_{\pi},m_{\pi})}.
\end{equation}
Here $W(\iota;E)$ is the sum of (A1.8) and (A1.9), and the
function $W(\iota \rightarrow \rho\rho;E)$ is 
\begin{equation}%A2.3
W(\iota \rightarrow \rho\rho;E) = \frac{1}{\pi}\,
\frac{E\,\Gamma(\iota \rightarrow \rho\rho;E)}{[M_{\iota}^2-E^2]^2+
[E\,\Gamma(\iota;E)]^2}.
\end{equation}
The partial width $\Gamma(\iota \rightarrow \rho\rho;E)$ is as
follows
\cite{Achasov}
\begin{eqnarray}%A2.4
\Gamma(\iota \rightarrow \rho\rho;E) & = & 
\frac{g^2_{\iota\rho\rho}}{8\pi}
\int^{E-2m_{\pi}}_{2m_{\pi}} 2E'dE'\:W(\rho;E')\:
\int^{E-E'}_{2m_{\pi}} 2E''dE''\:W(\rho;E'')\times 
\nonumber\\
&&\left[{\cal K}(E;E',E'')\right]^3\,[1-f(E;E',E'')] ,
\end{eqnarray}
where $f$ stands for the interference term, and the function
$W(\rho;E)$ is defined analogously to (A1.3), (A1.4).

In order to estimate $g^2_{\iota\rho\rho}/8\pi$ let us take into
account the experimental bound $\Gamma(\iota \rightarrow
\gamma\gamma)\times B(\iota \rightarrow K\bar K \pi) <  1.2$ kev
\cite{Behrend}. Since due to (35) $B(\iota \rightarrow K\bar K
\pi) > 0.4$, from this bound there follows $\Gamma(\iota
\rightarrow \gamma\gamma) < 3$ kev. Then, putting to use Ref.
\cite{Achasov} and the VMD model, one can obtain $\Gamma(\iota
\rightarrow \rho\rho) < 2$ MeV. From this bound, and, again,
with help of \cite{Achasov}, we obtain $g^2_{\iota\rho\rho}/8\pi
< 0.55 \,\mbox{GeV}^{-2}$. This result together with
(A2.2)--(A2.4) implies
\begin{equation}%A2.5
R < 10^{-3}.
\end{equation}

So, the $\iota \rightarrow \rho\rho$ contribution to the
annihilation $p\bar p \rightarrow \pi\pi\iota$ at rest is really
negligible. It is clear, that for $\iota \rightarrow
\omega\omega$ there should be the similar result. (Moreover, it
should be more strong since the width of the $\omega$ is much
less than the width of the $\rho$.)

\section*{Appendix 3}

\setcounter{equation}{0} 
\def\theequation{A3.\arabic{equation}}

Since the decay of $p\bar p$-atom to $\pi\pi\iota$ occurs near
the threshold, it is describable in the framework of $\chi$PT.
Let us build up the corresponding chiral effective Lagrangian.

With this purpose let us take into account the specific
properties of this decay \cite{OBELIX1}. The first property is
that the annihilation $p\bar p \rightarrow \pi\pi\iota$ at rest
is possible only from the iso\-singlet $^1S_0$ state of the
$p\bar p$-atom. So, its interpolating field, ${\cal P}^{\cal
N}$, must transform like $P^N$. The second property is that the
$\pi\pi$ system is produced in $S$-wave. Therefore, the pion
fields may contribute to the Lagrangian either without
derivatives or in combinations like
$\partial_{\mu}\,\pi\partial_{\mu}\pi$.  Finally, since the
relative angular momentum between the $\iota$ and the $\pi\pi$
system is also equal to zero, the interpolating field of $\iota$
and that of $\pi\pi$ should contribute without derivatives. The
above properties determine the following chiral-invariant
effective Lagrangian:
\begin{eqnarray}%A3.1
& L_{{\cal P}P\pi\pi}  = 
g^{}_1\,\langle u_{\mu}u_{\mu}\{{\cal P},P\} \rangle 
\, + \,\,
g^{}_2\,\langle [{\cal P},u_{\mu}][P,u_{\mu}]\rangle  &
\nonumber \\
& + \,\, g^{}_3\,\langle \chi_{+}\{{\cal P},P\}\rangle 
\,+ \,\,
g^{}_4\,\langle \chi_{-} {\cal P} \rangle + \,O(p^4). &
\end{eqnarray}
Here ${\cal P}$ stands for the baryon-antibaryon atom, $P$
stands for the nonet of the excited $q\bar q$ states and
$\phi^0$. (The glueball interpolating field does not contribute
to the Lagrangian due to the reasons discussed in Sections 2 and
4. Note, the latter property does not affect the final result.)
In what follows we consider ${\cal P} = {\cal P}^{\cal
N}\lambda^N/2$.

Putting $m_{\pi}^2 = 0$, one can show that only $\phi^0$
contributes to $p\bar p \rightarrow \pi\pi\iota$. (Generally
speaking, the excited state $P^N$ can contribute, too, but it
does not contribute to $\iota$.) After the superfluous terms are
rejected, in right hand size in (A3.1) there are remained
\begin{equation}%A3.2
\frac{g^{}_1}{2F^2}{\cal P}^{\cal N}\phi^0
\left(\partial_{\mu}\pi^{\!0}\partial_{\mu}\pi^{\!0} +
2\,\partial_{\mu}\pi^{\!+}\partial_{\mu}\pi^{\!-}\right) .
\end{equation}
From (A3.2) there follows the sought-for result
\begin{equation}%A3.3
L_{{\cal N}\iota\pi\pi} \: \propto \: {\cal P}^{\cal N}P^{\,\iota}
\left(\partial_{\mu}\pi^{\!0}\partial_{\mu}\pi^{\!0} +
2\,\partial_{\mu}\pi^{\!+}\partial_{\mu}\pi^{\!-}\right) .
\end{equation}

\end{document}